\documentclass[twocolumn,linenumbers,showpacs,showkeys,preprintnumbers,amsmath,amssymb]{revtex4} 

\usepackage{graphicx}
\usepackage{dcolumn}
\usepackage{bm}
\usepackage[T1]{fontenc}      
\usepackage{xspace}
\usepackage{pslatex} 
\usepackage{ifpdf} 

\newcommand{\al}{\ensuremath{\alpha}\xspace}
\newcommand{\ten}[1]{\ensuremath{\cdot 10^{#1}}}

\begin{document}


\title{Numerical Solution-Space Analysis of Satisfiability Problems}

\author{Alexander Mann}%
 \email{alexander.mann@physik.uni-goettingen.de}
 \affiliation{{II.} Institute of Physics, University of G\"ottingen, 
Friedrich-Hund-Platz 1, 37077 G\"ottingen, Germany}%
\author{A.K. Hartmann }%
\email{a.hartmann@uni-oldenburg.de}
\affiliation{Institute of Physics, University of Oldenburg,
Carl von Ossietzky Stra\ss{}e 9-11, 26129 Oldenburg, Germany}%

\date{\today}

\begin{abstract}
The solution-space structure of the 3-Satisfiability Problem (3-SAT)
is studied as a function of the control parameter $\alpha$
(ratio of number of clauses to the number of variables)
using numerical simulations. For this purpose, one has to sample
the solution space with uniform weight. It is shown here that standard
stochastic local-search (SLS) algorithms like ``ASAT'' and ``MCMCMC'' (also
known as ``parallel tempering'') exhibit a sampling bias.
 Nevertheless, unbiased samples of solutions can be obtained using the
``ballistic-networking approach'', which is introduced here.
It is a generalization of ``ballistic search'' methods and  yields also 
a cluster structure of the solution space.

As application, solutions of 3-SAT instances
are generated using ASAT plus ballistic networking.
 The numerical results are compatible with a previous analytic prediction
of a simple solution-space structure for small values of $\al$
and a transition to a clustered
phase at $\alpha_c\approx 3.86$, where the solution space breaks up into
several non-negligible clusters. Furthermore, in the 
thermodynamic limit there are, 
for values of $\alpha$ close to the SAT-UNSAT transition
$\alpha_s\approx 4.267$, always clusters without any frozen variables.
This may explain why some SLS algorithms are able to solve very 
large 3-SAT instances close to the SAT-UNSAT transition.
\end{abstract}


\maketitle



\section{Introduction}

The application of notions, analytical approaches and numerical
algorithms from statistical mechanics has lead to a better understanding
\cite{Mertens2002a,HA-W,MezardCSP} of NP-hard optimization problems 
\cite{Garey1979,Papadimitriou1994}. One main underlying question is, why
these optimization problems are computationally hard. 
This means no fast algorithms
are available, where the running times increase only polynomially
with the problem size.
The progress of gaining insight into this phenomenon has been considerable in
particular for the \emph{typical-case complexity}, where ensembles
of random instances are studied as a function of control parameters.
These ensembles often exhibit phase transitions where changes
of the effective ``hardness'' of the problem can be observed.
Often, these transitions are connected to changes of the structure
of the solution space, comparable to energy landscapes in physics. 
In particular, one is interested in the question, how the change
of the solution-space structures has an influence on the performance
of exact and stochastic algorithms. For example, for the
vertex-cover problem, which is defined on graphs, a clustering
transition has been found analytically \cite{vc2000} and numerically 
\cite{Barthel2004,Hartmann2008}
when increasing the edge density of Erd\H{o}s-R\'{e}nyi random graphs. This transition
coincides with a change of the typical-case complexity from polynomial
to exponential \cite{bauer2001}. For other optimization 
problems, the situation is less clear,
as for the \emph{satisfiability problem} (SAT), which we study in this work.

As we will explain, exact enumeration of solutions works well in one region
of the phase diagram, close to the SAT-UNSAT phase transition (see below), 
whereas Monte
Carlo approaches perform well in the opposite part of the phase diagram, 
 away from the SAT-UNSAT
transitions. Unfortunately, the clustering transition is located right
between these extreme parts, hence numerically difficult to study. We
use a stochastic algorithm in combination with a correction of the sampling
bias introduced by the stochastic algorithm to study the clustering phenomena.

The outline of the paper is as follows. In the second section,
we give the necessary background on SAT and on clustering of solution
landscapes. In the third section, we briefly explain the algorithms we
use to sample solutions and show that they exhibit a bias. Next,
we introduce ballistic networking and related methods, which we use
to correct for the bias. In section five, we show the results we
have obtained for random 3-SAT. Finally, we provide a conclusion and an
outlook.

\section{Background}

\subsection{Satisfiability}
Satisfiability  is one of the fundamental problems of computer
science, and has attracted a lot of attention over the past years,
also by physicists, due to its similarity to spin-glass problems.
It is the first problem proven to belong to the class of NP-complete problems 
\cite{Cook1971},
a class of problems for which no algorithm has been found yet that 
exhibits a polynomial worst-case running time as a function of the
problem size.
Therefore it is still a challenge to find algorithms which perform well 
on typical instances and to understand the underlying structure of the 
solution space which may hinder the performance of algorithms.

Satisfiability belongs to the class of constraint satisfaction
problems \cite{DIMACSh1997}: Given $N$ Boolean variables $x_i=0,1$ and
a Boolean formula $F$ describing a set of constraints, each of which
forbids a certain assignment of values to some of the variables, you are to
decide whether $F$ can be satisfied, i.~e., whether there is an
assignment $\vec x=(x_1,\ldots,x_N)$ such that all constraints are fulfilled
simultaneously.  In the $K$-SAT formulation, $F$ is given in
conjunctive normal form,
$$F = \bigwedge\limits_{m=1}^M (l^{m}_{1}\vee l^{m}_2 \vee \ldots \vee 
l^{m}_K)\,,$$
which describes a logical conjunction of $M$ constraints
(clauses) $C_m$ each containing a disjunction of $K$ literals 
$l^m_k$ ($m=1,\ldots,M\,;k=1,\ldots,K$) which are
either a variable $x_i$  or a logically
negated variable $\overline{x_i}$.

A certain assignment of values to the variables is called a configuration in the following. If a configuration satisfies all clauses in $F$ it is called a solution.
In the random $K$-SAT ensemble each clause is chosen randomly and uniformly amongst the $2^K {N \choose K}$ possible combinations in which no variable appears twice.

One defines a control parameter $\al = M/N$ which is the number of
clauses $M$ divided by the number of variables $N$.  For low \al the
problem is typically satisfiable whereas for high values of \al there
typically is no solution \cite{mitchell1992,selman1994}.  It has been
proven rigorously \cite{Friedgut1999} that the transition between the
satisfiable and the unsatisfiable phase becomes sharp for
$N\to\infty$. 
Whilst the position of the threshold for $K=2$ is known
exactly \cite{Goerdt1992}, for larger $K$ there are only numerical
estimates. In this paper we will stick to the $K=3$ case, where every
clause contains exactly three literals. The satisfiability transition
is located in this case at $\al_s = 4.267$ \cite{Mertens2006}.

\subsection{Cluster phenomena}

\label{sec:cluster_phenomena}
In addition to the SAT-UNSAT transition, analytical calculations
\cite{Mezard2005,Krzakala2007CSP} give rise to evidence that  there are further
(``structural'') phase transitions which refer to the formation of
disconnected clusters of solutions for high values of the control
parameter $\al$ in the
satisfiable phase. Formally, clusters in constraint satisfaction
problems can be defined as extremal Gibbs measures which gives the
following picture for Satisfiability:
For small values of \al all solutions are contained in one
connected component (cluster). When \al grows, more and more solutions
disappear so that at some point the cluster decomposes into smaller
clusters which initially, up to a threshold $\al_d$, make up only an
exponentially small fraction of all solutions, whereas above
$\al_d$ many clusters contribute to the statisical behavior.
Above a higher critical
value $\al_c$ we enter another type of clustered phase which is
dominated by a small number of large clusters.  The case of 3-SAT is
special, as here $\al_d = \al_c$, i.~e., we directly enter the phase
dominated by few clusters. The position of the dynamical threshold to
the clustered phase is predicted to be at $\al_c \approx 3.86$ 
\cite{Krzakala2007CSP}. 

This value is compatible with recent 
numerical results \cite{Zhou2009}, where the cluster structure was
investigated using the detection of community structures.
Unfortunately, the sampling was performed
using an algorithm, which does not exhibit uniform sampling of the solutions,
see below.
Anyway, there is no general rule how to translate the formal definition 
of clusters, which holds in the thermodynamic limit, 
to finite system sizes, hence other approaches than community
structures are possible.
For numerical studies often a very appealing
 approach is used, where a cluster is defined as the connected components 
in a graph where each solution is represented by a vertex and edges 
connect solutions differing in only one variable. 
This definition of a cluster will be used in this work as well.
For every two solutions belonging to the same cluster there is
therefore a ``path'' of configurations which all solve the SAT
instance  at hand. Unfortunately, this path can be long and peppered 
with many dead ends or loops which makes it very difficult to
decide  whether two configurations belong to the same cluster.  The
main problem when discussing clusters in high-dimensional
discrete solution spaces like that of Satisfiability is that one is
tempted to think of clusters as blob-like, well-seperated and
homogeneous structures in configuration phase like, e.~g., nano-clusters
formed by agglomeration of atoms. The clusters which occur in
high-dimensional discrete solution spaces are yet of a completely
different nature in that they are more like fragmented and interweaved
structures with lots of dead-ends, loops and holes which makes it difficult
to speak of spatially seperated clusters.

The existence of a clustered phase has been proven rigorously for
$K\geq 8$ \cite{Mezard2005}. In the language of statistical physics
this clustering corresponds to one-step replica-symmetry breaking
(1-RSB) \cite{Mezard2002Z,Mezard2002}. A further substructure in terms
of another clustering of solutions taken from one cluster, giving a
hierarchical structure of clusters is suspected where 1-RSB becomes
instable and higher steps of replica symmetry breaking occur
\cite{Montanari2004}.

What makes cluster phenomena interesting from the algorithmic point of
view is the question if (and if so in what way) clustering has an
influence on the performance of local search heuristics. 
Usually it is assumed, that the existence of many clusters
is an indication for a complicated ``rugged'' energy landscape,
which then also gives rise to many local minima, hindering the
performance of local search heuristics \cite{Montanari2004}.
In the same way, but with a
slightly different focus, Krz\k{a}ka{\l}a 
et al. \cite{Krzakala2007CSP}
propose that the appearance of locally frozen variables in
clusters is responsible for the  slowdown of heuristic algorithms close
to the SAT-UNSAT threshold.
A locally frozen
variable is a variable which takes the same value over all solutions
belonging to one cluster. A cluster containing at least one
frozen variable is
called frozen. One defines the freezing transition $\alpha_f$ as the smallest 
value of \al above which all solutions belong to frozen clusters.

To clarify the influence of phase transitions on the 
average computational hardness one can study the
performance of stochastic algorithms as a function of the 
control parameter \al. Of particular interest is the
algorithm-dependent value of \al up to which an algorithm shows linear-time
performance, and compare this to threshold values of \al
\cite{Ardelius2006}. 
Studies of stochastic
algorithms such as ASAT \cite{Ardelius2006}, WalkSat \cite{Seitz2005}
and ChainSat \cite{Alava2008} have shown
however that those algorithms have linear behaviour up to values
considerably beyond the clustering transition. This suggests
that the cluster transition has no impact on the performance of local
search algorithms as long as there are precautions against entropic
traps. 
It is remarkable that ChainSat has this behaviour
 although it is greedy ``in a weak sense'' as it never allows steps
which increase the number of unsatisfied clauses. Naively one would
therefore expect it to get trapped very easily in local minima. 
The authors of \cite{Alava2008} interpret this as evidence 
for the belief that true local
minima are very rare in high-dimensional search spaces. These
results could also indicate that indeed it is more the (non-) existence
of frozen clusters, which is responsible for the performance of 
local-search algorithms.


Nevertheless, for small instances there are always some frozen variables. 
Therefore in \cite{Ardelius2008} a different notion of frozen 
clusters via the \emph{whitening core}
is used. There one looks, for each solution,
 interatively for variables which can be flipped since they appear only
in clauses satisfied by other variables or which contain variables already
detected in the whitening core.
The position of this  freezing transition was then 
calculated by exact enumeration and clustering of all solutions 
for sufficiently small system sizes and is expected to lie at 
$\al_f = 4.254$ close to but below the satisfiability transition. 
In a similar way, 
\cite{Ardelius2007} finds a cluster condensation transition in the 
solutions generated by ASAT very close to $\al_s$, again these
results rely on a non-uniform sampling of the solutions.
Anyway, these results
are compatible with the observation of a 
good performance of local-search algorithms
close to the threshold $\al_s$.

\subsection{Algorithmic treatment of SAT}
Algorithms for SAT include a broad spectrum, both stochastic and exact, from simple and straight-forward algorithms like RandomWalksat \cite{Papadimitriou1991} and WalkSat \cite{Selman1996,Kautz1997} 
to complex algorithms like DPLL \cite{Davis1962} 
and message-passing algorithms such as Belief Propagation (BP) and Survey Propagation (SP) \cite{Mezard2002}.
For small systems exact enumeration of all solutions is possible using one of the numerous standard algorithms \cite{gkss08:sat-kr,bayardo2000} 
such as the aforementioned DPLL.
It can be shown \cite{cocco2001}
that deterministic algorithms have longest average run times close to 
$\al_s$ reflecting the difficulty of deciding whether a given SAT formula 
is satisfiable or not. 
The problem with exact enumeration is that it is limited to small systems due to hardware restrictions, especially because of the memory needed to store the huge number of configurations, as the number of solutions grows exponentially with system size.
Furthermore the number of solutions is not a continuous function when crossing the satisfiability threshold, but it drops from a finite number to zero. This corresponds to a non-zero entropy at the phase transition, the entropy per variable grows approximately linearly with decreasing $\al$ \cite{Monasson1996,Biroli2000}.
In turn this means that even very close to the satisfiability threshold the number of solution grows exponentially, and quickly becomes so large that it is not feasible to enumerate all solutions even in this regime.
From counting all solutions using DPLL for systems up to $N=160$ we can estimate the solution entropy $s = S/N$ near the phase boundary $\al = 4.25$ to be roughly between $0.11$ and $0.12$ 
which gives more than $10^8$ solutions already for $N>160$, thus taking up at least 
2 GB.

To overcome these limitations one can turn to stochastic algorithms which, starting at an arbitrary configuration, do successive changes either completely randomly or based on a heuristic evaluating information about the local configurational neighbourhood. Stochastic algorithms are not guaranteed to find a solution, even if solutions do exist, but they can be significantly faster than deterministic algorithms.
It is thus possible to obtain solutions for much larger systems, but on the other hand stochastic algorithms can never prove that there is no solution, tests for unsolvability can only be done by using deterministic algorithms.

In this paper we study the cluster structure numerically for $K=3$, which requires unbiased sampling of the solution space. Different types of sampling algorithms are studied and shown to be biased. We therefore present an algorithm which uses a different approach to create a survey of the cluster structure of Satisfiability instances from which it is then possible to derive unbiased samples.
It is an improvement on the ``Ballistic Search'' algorithm which has originally been applied to spin glasses \cite{Hartmann2000a,Hartmann2000b,Hed2001}. 
The main advantage of the algorithm is that it is able to provide an overview of the cluster structure of the solution space without having to enumerate all solutions which is no longer possible already for moderate numbers of variables.


%
%

\section{Sampling algorithms}
\label{sec:bias_in_standard_sampling}
\subsection{Bias in SLS algorithms}
\label{sec:biasASAT}
If stochastic local search (SLS) algorithms found all solutions with the same probability one could use them directly to probe the solution space. Unfortunately, this is not the case as we will see in this section.
Later on in this article we will use ASAT as solution generator so we use it here for an exemplary presentation of the bias in SLS algorithms.

ASAT is a simplified variant of Focused Metropolis
 Search \cite{Seitz2005} and was first described in 2006
 \cite{Ardelius2006}.  It starts at a random configuration and in each
 step picks a variable from an unsatisfied clause. This variable is
 flipped if either this decreases the number of unsatisfied clauses,
 or otherwise with a constant probability $p$ which is a tuning parameter
 of the algorithm.  ASAT has run times linear in the system size at
 least up to $\al = 4.21$ on 3-SAT. For instances of moderate size
 like those we study here, it can well be used beyond this point
 \cite{Ardelius2007}. 

The test procedure is very simple: For a randomly chosen 
small instance we run ASAT again and again starting each time from a different randomly chosen configuration and count how often each solution returned by ASAT is found. If there were no bias we would expect the histogram of solution multiplicities to be flat except for statistical fluctuations around a plateau value, i.~e. the histogram should resemble a Gaussian error function. 

\begin{figure}
  \includegraphics[angle=-90,width=\linewidth]{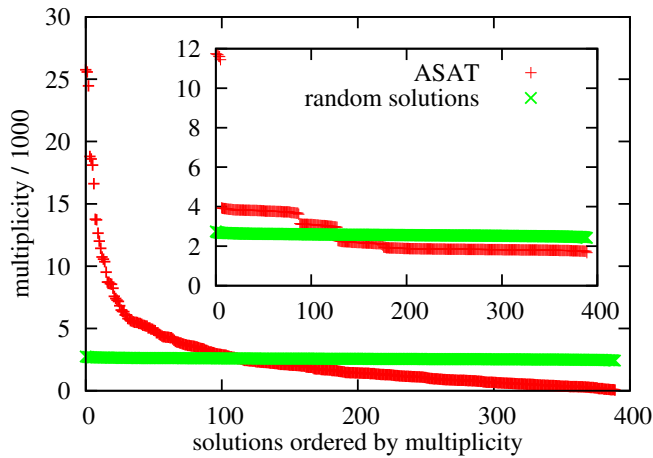}
  \caption{
    Multiplicities of solutions found by ASAT in $10^6$ runs for a randomly chosen instance with $N=50$, $\al=4.0$, compared to an unbiased distribution.
    Inset: Multiplicities of the ASAT solutions after an additional $T=0$-MC step with 10 sweeps. 
  }
  \label{fig:ASATmultiplicity}
\end{figure}

Fig. \ref{fig:ASATmultiplicity} shows the resulting histogram, in
comparison to a histogram filled with the same number of random
integers drawn from a truely flat distribution over the range
corresponding to the number of solutions of the SAT instance
showing what the distribution should look like if there were no bias.
Clearly there is a strong bias favouring some solutions over others.
To quantify the deviations we use a $\chi^2$ test, and calculate the
$p$-value giving the probability that an unbiased sampling process
yielded a sample deviating at least as much than the one at hand. The
$p$-values numerically are smaller than $10^{-323}$ 
(i.~e., the resolution of our \texttt{double} numbers).

To test whether this bias can be corrected in a simple
way, we did a further check, where instead of using the solutions
returned by ASAT directly, for each solution found by ASAT a solution
from the same cluster was generated using a $T=0$ Monte Carlo (MC)
search starting at the ASAT result.
The outcome of this modification is shown in the inset of 
Fig.~\ref{fig:ASATmultiplicity}
for the same SAT instance as before. The
distribution now clearly has 5 plateaus corresponding to the 5
clusters of the solution space and looks much flatter 
but exhibits still some bias. One sees that 
most of the ASAT solutions stem from
the smallest cluster, hence the sampling does not respect the cluster
size. Hence, ths bias can be strongly decreased by additional
$T=0$ MC simulations, but not completely.
Further checks showed that the bias persists independently of the system 
size. 

Since we want to study clustering properties of the solution
ensemble we need to
remove the bias completely and 
 sample solutions in proportion to the cluster sizes. 
To ensure this, we will perform
 reweighting using the Ballistic-Search algorithm as described in
 section \ref{sec:ballistic_search_intro}. Before we come to the
 Ballistic Search, we will show in the next section that MCMCMC,
 another important sampling method, fails on  sampling SAT
solutions uniformly as well.

\subsection{Bias in MCMCMC}
The Metropolis-Coupled Markov Chain Monte Carlo (MCMCMC) method, first
proposed in 1991 by Geyer \cite{geyer1991}, also known as Parallel
Tempering \cite{hukushima1996,NewmanBarkema}, is a powerful and versatile
tool, commonly used in biophysics and statistical physics to perform
equilibrium simulations and to generate unbiased samples in large
configuration spaces.
%
%
MCMCMC uses a set of replicas of single instances, simulated in
parallel at different temperatures and linked by global updates in
which replicas are swapped pair-wise with an acceptance probability
depending on their energy difference and temperature spacing
(Metropolis-Hastings criterion),
thus facilitating the tunneling through barriers seperating local
minima of the phase space \cite{juanjo2003}.

To study the performance of MCMCMC on SAT we employ a histogram test
similar to the one described in section \ref{sec:biasASAT} for the
performance of ASAT.  For several values of $\al = 1.00 \dots 4.25$
scattered over the satisfiable phase, the number of variables $N$ is
chosen such that expected number of solutions is 1000. This, e.~g.,
results for the smallest value of \al considered here 
in a system size $N = 14$, 
while for the highest value of \al, $N =  50$ is feasible.

We apply a straight-forward implementation of MCMCMC to a set of 50
instances for each value of the control parameter \al,
where we use 15 temperatures, the lowest, at which the samples are
taken, being initially $T_0=0.1$, the highest such that the
corresponding energy is found to be approximately $M 2^{-K}$ which is
the expected energy of a completely random configuration. Every 1000
steps the temperatures are adjusted 
to drive the replica exchange rate between neighboring temperatures
towards 50~
lowest temperature fixed at $0.1$.  The procedure chosen to adjust the
temperatures leads to a distribution of temperatures where for the
lowest temperatures the exchange rates indeed reach 50~
whereas the highest temperatures all gather in the random phase.
This can be seen as an indication that the number of temperatures used is sufficient to allow the replicas to travel between the constrains, i.~e.
the highest temperatures are indeed located in the ``paramagnetic phase''.
We take one sample every second sweep to generate a total of $10^6$ samples. Only successful sampling steps are counted, i.~e. those where the energy of the configuration at $T_0$ is zero.

\begin{figure}
  \includegraphics[angle=-90,width=\linewidth]{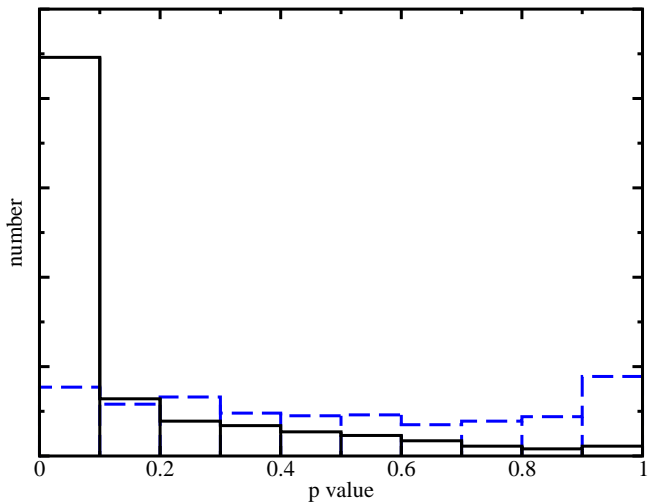}
  \caption{Bias of MCMCMC: Dependence of $p$-value on the number of clusters, integrated over all $N$ and \al. Dashed line: $p$-values for instances 
with only one cluster, solid line: $p$-values for instances 
with more than one cluster.}
  \label{fig:mcmcmc_bias_1} 
\end{figure}

The histograms with the resulting distribution look pretty close to
those drawn from a flat distribution (not shown here). We
again use the $p$-values  obtained from a $\chi^2$ test to quantify
the deviations and find that in most cases MCMCMC gives reasonably
flat distributions,  hence this method appears to exhibit on
the first sight a much lower sampling bias. Nevertheless,  there also
are a number of histograms having a significant bias exhibiting very
small $p$-values. The higher the value of the control
parameter \al is chosen, the larger becomes the spread of the
distribution of $p$-values  towards extremely small values. 
This can be seen from Fig.\
\ref{fig:mcmcmc_bias_1} where the distribution of $p$-values,
integrated over all system sizes and values of the control parameter,
is shown for the two cases where the instances exhibit one or more
than one cluster of solutions, respectively.  (The number of clusters
can easily be calculated exactly for these small instances.)  In the
case that the solution space consists of only one connected component,
the $p$-value distribution is flat showing that MCMCMC works unbiased
as expected.  The presence of clustering on the other hand leads to a
bias or imbalance in the sampling process resulting in a strong peak
at small $p$-values.  

Since we are interested in particular in those instances which
exhibit many clusters, MCMCMC turns out to be not suitable as well, since all
instances have to be sampled correctly. Note
that for larger system sizes, the number of instances having just one
cluster, where MCMCMC seems to work well, will strongly decrease. Hence,
for large system sizes, MCMCMC will exhibit a bias for basically all instances
of interest.
To create an unbiased sample we need a different method which will 
be presented in section \ref{sec:ballistic_search_intro}.

\section{Ballistic search}
\label{sec:ballistic_search_intro}
Here, as mentioned in Sec.\ \ref{sec:cluster_phenomena}, 
we are using the neighbor-based definition of clusters: two
solutions are considered to be in the same clusters, if there
exists a path in solution space consisting of single-variable flips.

We use Ballistic Search, which
 has been introduced in the year 2000 as a method for 
studying ground-state properties of spin glasses \cite{Hartmann2000a}. 
The approach
 is able to provide a survey of the cluster landscape using
stochastic algorithms, in particular without the need to enumerate all
ground states as it is usually necessary when one aims at
clustering.  The sheer number of ground states forbids exact
enumeration when studying spin glasses, and, as mentioned above, the same
holds for SAT.  We therefore use this method which relies
on generating a survey of the most important clusters.

The survey consists of a set $A=\{A_i\}$ where each element $A_i = (\{c^{(i)}_j\},\Sigma^{(i)})$ represents one cluster and consists of a (small) set of solutions $\{c^{(i)}_j\}$ from the cluster $i$ and an estimate $\Sigma^{(i)}$ of the size of cluster $i$. The survey should cover all clusters, or at least all but those which are negligibly small. One can then sample the whole solution space with correct weights by generating the desired number of solution samples from the representative sets of solutions for each cluster according to the respective cluster sizes.

Two main ingredients form the basis of the Ballistic Search algorithm.
Firstly, the above described data structure storing small sets of 
representative solutions for each cluster instead all solutions.
Secondly, a ``ballistic path search'' is used to 
analyse the cluster space and generate the survey from a given 
set of solutions.
The basic operation of this procedure 
is that we have
to determine for any given pair $c_a$, $c_b$ 
of solutions, whether they belong to
the same cluster. This has to work
 under the assumption that for the case of $c_a$, $c_b$ belonging
to the same cluster,
a complete nearest-neighbor path of solutions 
between $c_a$ and $c_b$ is \emph{not}
contained in the set of already found
solutions. Instead,
one searches stochastically for paths between $c_a$ and $c_b$ by starting at one solution and subsequently changing a randomly chosen \emph{free} variable.
A variable is called \emph{free} if its value can be changed without violating any constraint, so that one never leaves the solution cluster.
This is repeated until either the target solution is reached or no free variable is left, because as an additional stop condition every variable shall be touched at most once. (Because of this additional constraint the path search is called ``ballistic''.) This implies that in a successful ballistic path search the number of steps taken is always equal to the Hamming distance of the solutions, 
i.~e., the number of variables in which the two solutions differ.

\begin{figure}
  \includegraphics[width=0.6\linewidth]{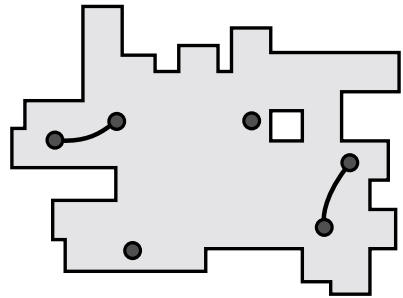}
  \caption{Between those 6 solutions (black circles) from the same cluster (light grey) Ballistic Search has found only 2 paths. The apparent number of clusters is 4. We need to increase the density of solutions to make Ballistic Search more efficient.}
  \label{fig:explanation_bs1}
\end{figure}

\begin{figure}
  \includegraphics[width=0.6\linewidth]{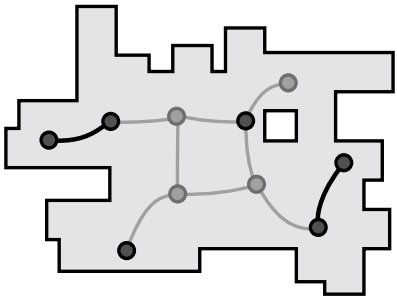}
  \caption{Adding solutions (grey) has yielded a correct identification of the cluster, because more paths (grey) have been 
found, now connecting all solutions.}
  \label{fig:explanation_bs2}
\end{figure}

Figures \ref{fig:explanation_bs1} and \ref{fig:explanation_bs2} give a
graphical description of how the Ballistic Search algorithm works.  We
start with some randomly generated solutions depicted as black circles
in Fig. \ref{fig:explanation_bs1}, all of which belong to the same
cluster which is drawn in grey in a 2-dimensional cartoon of the
N-dimensional configuration space. For the sake of simplicity we
assume here that all solutions belong to the same cluster, the
generalization to more than one cluster is obvious.  Running the
ballistic path search we find that some of the solutions can be
connected by paths drawn as lines in the picture, i.~e., 
for these solutions the algorithms correctly finds  that
they belong to the same cluster. The problem is that for low solution
densities the average distance between solutions is large, and the
efficiency of the ballistic-path search strongly decreases with larger
distances  
\cite{Hartmann2000a}. Therefore we only find a few paths and
the apparent number of clusters in this example is larger than 
its true value, cf. Fig. \ref{fig:explanation_bs1}.
What we need to do is to increase the number of solutions by
rerunning ASAT. 
For few added solutions, the measured number
of clusters will increase, since
only few additional paths within clusters are detected, less than the number
of added solutions.
When generating even more solutions,
 we will find that the apparent number of
clusters at some point no longer increases, but instead it decreases
as more and more paths between solutions are found, until finally all
solutions are correctly assigned to the same cluster as shown in
Fig. \ref{fig:explanation_bs2}.

\subsection{Ballistic networking}

\begin{figure}
  \includegraphics[width=0.6\linewidth]{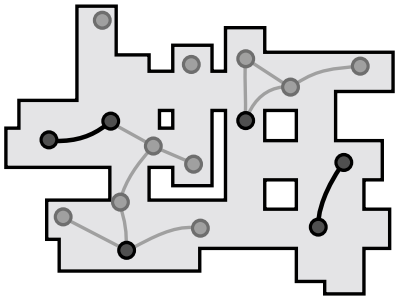}
  \caption{This cluster has a more complex structure than the one in Fig. \ref{fig:explanation_bs2}, illustrated by the additional holes. Here adding even more new solutions (grey) than before does not work. Still not all solutions are recognized as belonging to the same cluster.}
  \label{fig:explanation_bs3}
\end{figure}

Our studies have shown that the simple ballistic-path search algorithm has very low efficiency when applied to Satisfiability which can be attributed to a 
high complexity of the solutions space or large sizes of the clusters. 
We therefore developed a refinement of the algorithm named ``ballistic networking'', which is a very general extension of the ballistic path search so that it can readily be applied to other problems.

\begin{figure}
  \includegraphics[width=0.6\linewidth]{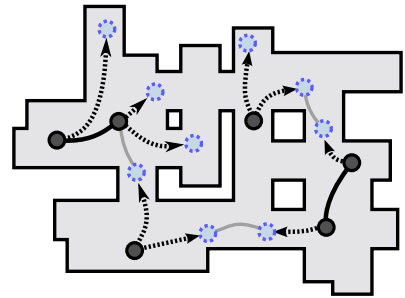}
  \caption{Ballistic Networking improves on the result of Ballistic 
Search by not adding solutions randomly, but adding solutions from the 
same cluster using a $T=0$ MC search (arrows). Now all solutions are 
found to belong to one cluster.
  \label{fig:explanation_bs4}}
\end{figure}

The idea of the algorithmic refinement is to increase the probability
of identifying two solutions, origin $c_a$ and target $c_b$,
 as belonging to the same cluster using
ballistic path search by again increasing the number of solutions. 
Instead of using ASAT to generate more solutions we generate 
$2n_\text{add}$ additional solutions by performing independent $T=0$
MC simulations starting at $c_a$ and $c_b$, respectively.
Hence, we are sure that the additional solutions belong to
the same cluster as their respective ``parent'' solution.  We then
try to find connections using the ordinary ballistic path search
between all $(n_\text{add}+1)^2$ pairs of solutions, where 
one solution belongs to
the origin and one to the target. If at least one path is found,
it is clear that $c_a$ and $c_b$ belong to the same cluster. We
 apply this test to all pairs of solutions which have not yet been
found to belong to the same cluster.  An artist's view of this
improvement is given in figures \ref{fig:explanation_bs3} and
\ref{fig:explanation_bs4}. Fig. \ref{fig:explanation_bs3} shows how
the standard ballistic search fails due to a more complex structure of
the cluster, although even more solutions than before have been
used. In Fig. \ref{fig:explanation_bs4} the solutions found by the
$T=0$ MC search are drawn as points connected to their parent solution
by arrows. We can see that the number of successful ballistic path
searches (gray lines) does not have to be very high, but still is
enough to correctly identify the cluster.

\begin{figure}
  \includegraphics[angle=-90,width=\linewidth]{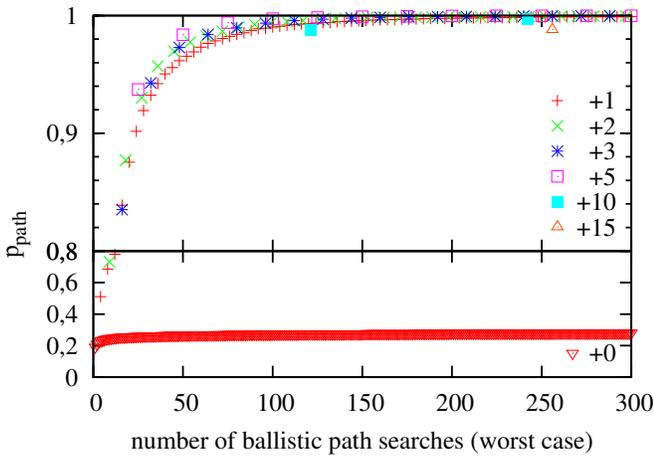}
  \caption{Comparison of ballistic path search without and with additional solutions (``ballistic networking'') for $N=128$ and $\al = 3.0$. We show the probability for finding a path between two solutions generated from the same cluster, as function of the total number of ballistic path searches between all pairs of parent and children configurations averaged over 1000 runs. The case ``+0'' corresponds to the original ballistic path search.}
  \label{fig:compare_BS_BN}
\end{figure}
Indeed this procedure improves the performance of the search so much
that it outweighs the additional effort of having to carry out
$(n_\text{add}+1)^2$ ballistic path searches instead of one.  Fig.
\ref{fig:compare_BS_BN} shows a comparison of the performance of
ballistic path search without and with additional solutions.  The case ``+0''
corresponds to the bare ballistic path search. The horizontal axis
shows the number of ballistic path searches which have to be carried
out in the worst case, i.~e., when no connecting path is found.  We
found $5 < n_\text{add} < 10$ to be a suitable range for the
system sizes under study. 

When creating the additional solutions to test whether a pair of
solutions $c_a$, $c_b$ belongs to the same cluster, to improve
the success probability, one can think of
introducing a bias into the $T=0$ MC search which pushes the
additional solutions derived from the first solution $c_a$ closer to
the second solution $c_b$ and vice versa.  Indeed we found that such a
bias has a positive influence on the success probability of the
ballistic path search. Yet we did not use this bias in the
implementation, because the positive influence comes at  a
high cost. For each pair of solutions to be tested a dedicated biased
set of additional solutions has to be generated which cannot be reused
when comparing either $c_a$ or $c_b$ to a third solution. The
necessary computational effort for generating each time new biased
configurations by far outweighs the positive effect of the bias.

\begin{figure}
  \includegraphics[angle=-90,width=\linewidth]{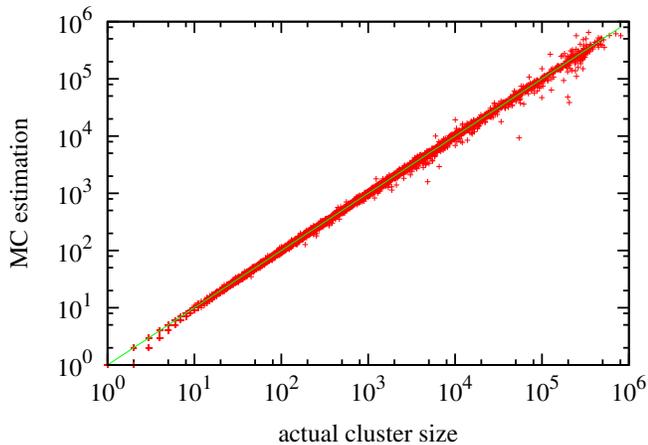}
  \caption{Comparison of the actual number of configurations per cluster to the number estimated by Monte Carlo integration. Each point corresponds to one cluster.}
  \label{fig:cluster_size_est}
\end{figure}

The second part of the cluster survey $A$ 
consists of the sizes $\Sigma^{(i)}$ of the clusters.
An exact calculation of the cluster size is possible, but takes too long, since it typically grows exponentially with $N$.
We have therefore examined several different estimation methods with
respect to their reliability in giving a correct estimate for the
cluster size by comparing the estimated cluster size to the exact
cluster size on a random ensemble of clusters for different values of
$\al$ and small system sizes $N$.  The best method known to us has
been found to be the estimation of the cluster size using a Monte
Carlo integration as it has been used in \cite{Hartmann2002} in an
application to spin glasses.
Fig. \ref{fig:cluster_size_est} shows a comparison of the actual
number of configurations in one cluster to the number as estimated by
Monte Carlo integration.  We used several combinations of $N$ and \al
where the total number of solutions (over all clusters) did not exceed
$5\ten6$, such that all clusters could be calculated exactly, and
afterwards for each cluster the MC estimation was run.

\subsection{Implementation}
\label{sec:implementation}
%
%
%
%
%
Combining ballistic networking and the cluster size estimation the full algorithm is comprised of two alternating steps.
The first step is to generate a given number (of the order of 1000) of solutions of the Satisfiability instance at hand using the ASAT algorithm.
The tuning parameter of ASAT is chosen to be $p=0.21$ 
which is the optimal value as given in \cite{Ardelius2006}.
In the second step Ballistic Networking of the solutions found by ASAT is done as described above to create the cluster survey, and then
 the sizes of the clusters are estimated.
Afterwards ASAT is run again and another set of new solutions is created. The cluster survey is then updated using Ballistic Networking on the new solutions and the solutions representing the so far found clusters in the existing survey. Here new clusters may be found, and if so, their size is estimated.
This is repeated until the cluster survey is considered complete, i.~e.,
 no more relevant clusters are found.

From the cluster survey for each instance a set $\mathfrak U$ of
unbiased solutions can be generated using the cluster-size
estimates. For each solution to be generated for a given instance,
first a cluster from the survey is selected with a probability
proportional to the cluster size. One solution is selected from the
set of representative solutions and starting from this solution a
$T=0$ MC search is performed finally giving the solution to be used in
the analysis.

Defining a good stopping criterion is a crucial point of
 the algorithm.  As the cluster number in Satisfiability can be rather
 large, we decided not to generate all clusters, but all except for
 those which contain only a neglegibly  number of solutions.  For this
 purpose we monitor the total cluster weight $\sum_i \Sigma^{(i)}$.
 We run the algorithm until the total cluster weight has not
 increased by more than $0.5~\%$ over the last half of solutions
 included in the clustering process.  We store the order in which the
 solutions have been generated by ASAT and label each cluster with a
 number telling the position of the earliest solution which has been
 found to belong to this cluster.

When trying to optimize the number of new solutions added in each round 
one has to consider two competing effects:
On the one hand adding solutions --- as in ordinary Ballistic Search --- may reveal that two clusters actually are parts of the same cluster, connected maybe by only a small path in configuration space which has been too hard to find with fewer solutions.
On the other hand increasing the number of solutions makes Ballistic Networking slower and, even worse, increases the probability of false new clusters 
which in turn can lead to an on-going increase of the cluster number and total cluster weight and thus to a failure of the stopping criterion. 

The system sizes which can be reached using the method described above depend, of course, on the control parameter $\al$.
For small \al all solutions are contained in only one large cluster where there are many possible paths between configurations so that Ballistic Search is very efficient and system sizes of a few hundred variables are possible.
For high \al values in the solvable phase, the number of solutions 
is small so that in this regime Ballistic Search still is rather efficient due to the small extent 
of the clusters and relatively large system can be done.

\begin{figure}
  \includegraphics[angle=-90,width=\linewidth]{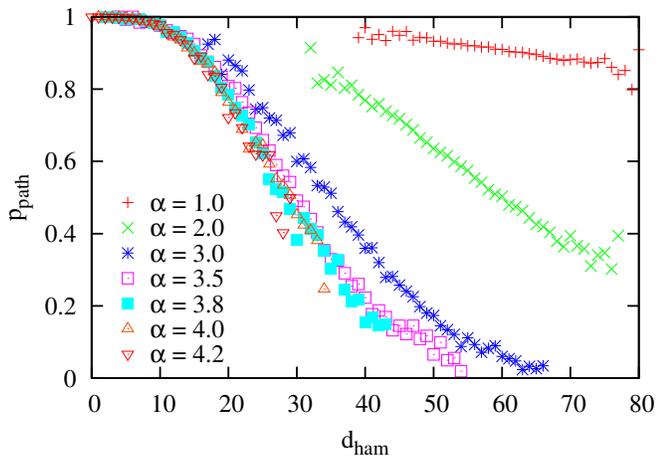}
  \caption{Dependence of $p_\text{path}$ of ballistic path search on \al and $d_\text{ham}$ for $N=128$. As the typical distance of two solutions depends on \al, so do the ranges shown in the plot.
For small distances 
$d_\text{ham}\le 15$, 
a path is usually found, while for large distances, the
probability depends on the value of the control parameter $\al$, 
which can be understood
by a more and more complex structure of the solution clusters.}
  \label{fig:BS_alpha_dep}
\end{figure}
Fig. \ref{fig:BS_alpha_dep} shows the dependence of the success
probability of ballistic path search on the Hamming distance
$d_\text{ham}$ between the configurations  for $N=128$, for
different values of the control parameter \al. Up to $\al=3.8$ the
probability decreases strongly with increasing \al, as the clusters
develop more holes. Above this point the curve is approximately
independent of \al.  We also find that the probability decreases
weakly with increasing system size (not shown).   The fact that the
average distance between solutions decreases with \al makes Ballistic
Networking most difficult in the intermediate regime around $\al
\approx 3.3$.   Here the number of ballistic path searchs needed to
find a connection between two solutions from the same cluster is
highest. The cluster structure seems to be such that there are many
``dead ends'' in which the search may get stuck. Together with the
high number of clusters, which enters quadratically in the
 running time,
this limits the reachable system size.  All in all, Satisfiability
instances of up to $N=144$ variables were doable in reasonable time
over the whole range of interest $3.0 < \al < 4.2$ while for smaller
intervals of the control parameter, we also studied $N=256$.


\section{Results}
\label{sec:results}

Here, we study the behavior of random 3-SAT instances
as a function of the parameter \al.
This is meant in the sense 
that we generate an instance using a given number $N$ of variables
and a  set of (arbitrarily ordered) clauses $C_m$ ($m=1,\ldots,M_{\max}$). 
We chose $M_{\max}=\al_{\max} N$,
where $\al_{\max}$ is the largest value of the control parameter we want to
consider. We can study the behavior
of each instance as function of $M\le M_{\max}$ by considerung each time
exactly the clauses $C_m$ for $m=1,\ldots,M$. Also, we can average over
these distances for each value of the control parameter.

\subsection{Hierarchical clustering}

For the analysis \cite{practical-guide2009}  of the behavior
of 3-SAT as a function of the control parameter $\al$, 
we start by looking at the hierarchical structure of a set $\mathfrak U$ of 
solutions  sampled for a typical 3-SAT instance.  
We have
used ``Ward's algorithm'' \cite{Ward1963, JainDubes}, an agglomerative
hierarchical matrix updating algorithm, on the set $\mathfrak U$ to
extract a hierarchical clustering from which we can then draw a visual
representation of the solution space.

Ward's algorithm has been applied in many different fields ranging
from RNA secondary structures over optimization problems to spin
glasses \cite{Higgs1996, Hed2001, Hartmann2008, Zhou2009}.  It is an
iterative procedure where initially each configuration comprises a
single item cluster.  In each step those two clusters are merged which
have minimal distance with respect to an effective distance measure
chosen such that the sum of the variances in each cluster is
minimized.  After each merger, the distances of the 
remaining clusters to the new cluster have to be calculated, for details
see, e.~g., Ref. \cite{JainDubes}. Finally, one
reorders the configurations according
to the hierarchy obtained in the iterative merging process, and draws a
color-coded visualization of the distance matrix.

\begin{figure}[ht]
  \centering
  \begin{minipage}[b]{0.27\textwidth}
       \includegraphics[width=\textwidth]{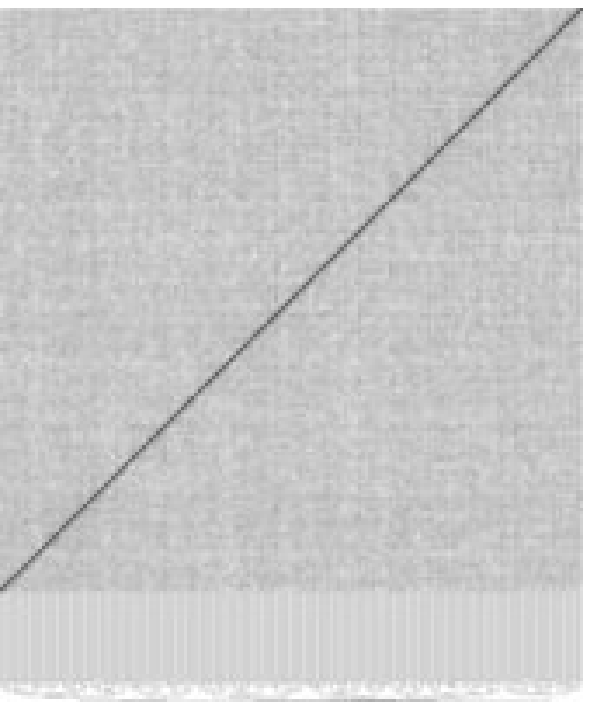}
       \includegraphics[width=\textwidth]{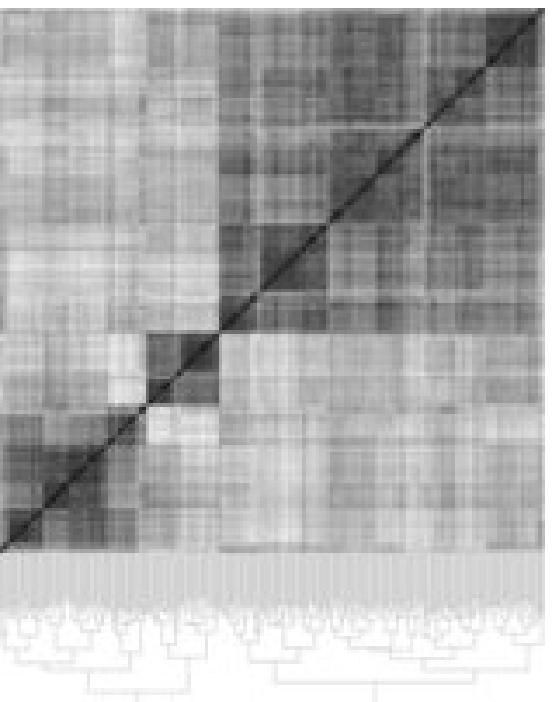}
       \includegraphics[width=\textwidth]{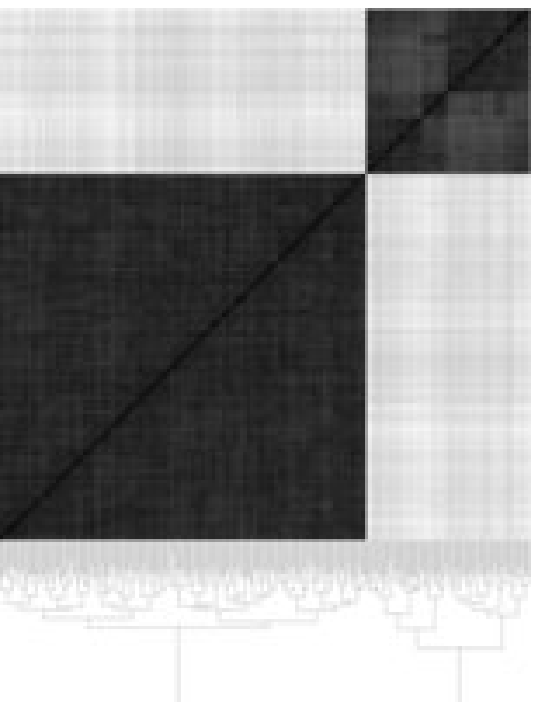}
  \end{minipage}
  \caption{
    The hierarchical structure resulting from Ward's algorithm visualized 
both as tree structure (dendrogram) and distance matrix, for $N=256$ 
and $\al = 1.00$ (top), $4.00$ and $4.25$ (bottom). 
Darker grey scales correspond to smaller distances.
  \label{fig:theMatrix}
  }
\end{figure}

Next, we present some results for a typical instance. We chose one 
which exhibits its SAT-UNSAT transition close to the 
numerical estimate of the ensemble average $\alpha_s = 4.267$ 
given in \cite{Mertens2006}.
Fig. \ref{fig:theMatrix} shows the color-coded distance matrices and 
the dendrogram which were generated for three different values of $\al$ . 
The difference in the solution landscape and cluster structure between 
the phases is clearly visible.
For low \al the Ward matrix is featureless and homogeneously grey. All 
solutions belong to one single cluster and the phase space shows no 
specific features.
In the intermediate range one sees box-like structures along the
diagonal in a darker grey. These correspond to clusters, because
darker means smaller Hamming distance and the solutions inside a
cluster are closer to each other than to other solutions.  Some of
these boxes show a substructure which can be interpreted as the
solutions from this cluster themselves forming sub-clusters. This is
consistent with the theoretical prediction of replica-symmetry
breaking beyond 1-RSB in the intermediate \al range
\cite{Montanari2004}. Nevertheless, as mentioned in the introduction,
it is to be expected that most of the clusters are not relevant in the 
thermodynamic limit and a small number of clusters contains almost 
all solutions. 
For higher values of \al the substructures inside the
clusters become washed out whereas the first-level cluster structure
becomes more pronounced as the cluster become smaller.  In the replica
symmetry breaking framework this would be interpreted as a vanishing
of higher level RSB above a certain threshold, but this cannot be
deduced from looking at single instances, of course.

%
%

\subsection{Averaged quantities}

%
%
%

%
%
\begin{figure}
  \includegraphics[angle=-90,width=\linewidth]{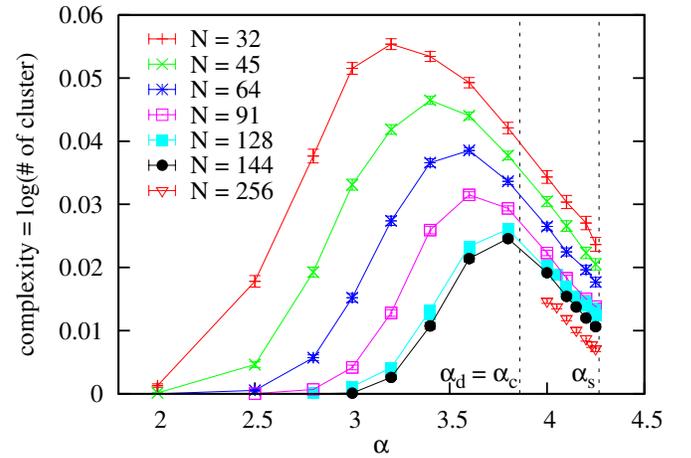}
  \caption{Complexity $c$ as a function of $\al$ and for several system sizes between $32$ and $144$. Lines have been drawn to guide the eye. The error bars give statistical errors.}
  \label{fig:complexity}
\end{figure}
The complexity $c = \frac 1N \log N_c$  
is defined as the logarithm of the number of clusters normalized to the system size.
Fig. \ref{fig:complexity} shows the complexity as a function of $\al$ 
and for system sizes up to $N=256$ variables averaged over 
200--500 instances for each value of \al and each value of $N$. 
The number of clusters was taken directly from the cluster surveys 
created using the Ballistic Networking method described above.

For the ``easy'' part of the satisfiable phase,
where the value of the control parameter \al is small,
 there is only one cluster, thus the complexity is zero.
In an intermediate range the number of clusters grows peaking at a 
value which is strongly affected by finite-size effects and then 
becomes smaller again.
This behaviour reflects the theoretical prediction of one single cluster in the low \al regime ``crumbling'' into smaller pieces when \al is increased and the clustered phase is reached. For even higher \al the vanishing of solutions leads to the disappearance of clusters 
and the cluster number decreases again. Clearly,
 the peak of the complexity curves seems to converge towards 
$\al_c$ with increasing system size, in accordance with the analytic
predictions \cite{Krzakala2007CSP}.

Looking at this plot one has to keep in mind that the stopping criterion used in the algorithm is based on the number of solutions covered by the clusters that were found so far. In a phase with a large number of small clusters we will miss small clusters if they only comprise a negligible part of the solutions (in the sense of the stopping criterion) and therefore underestimate the number of clusters. It is therefore natural that the complexity found here is lower than the one given in \cite{Ardelius2008}.
After all the complexity shown in the graph is only a lower bound for the true complexity respecting all clusters.
In phases dominated by few and large clusters its value should nevertheless be close to the true value.

\begin{figure}
 \includegraphics[angle=-90,width=\linewidth]{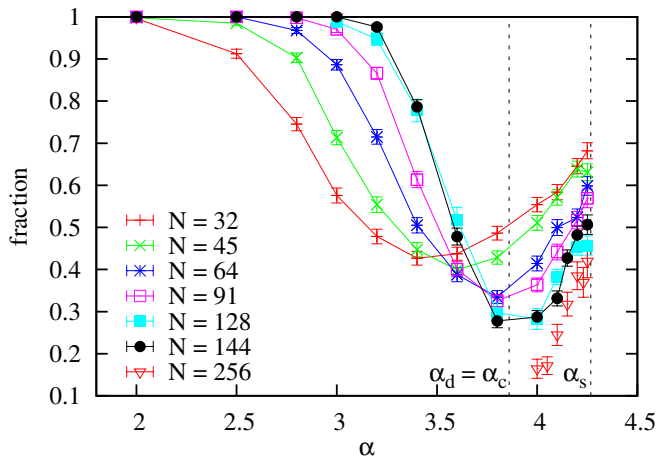}
\caption{Fraction of the weight of the largest cluster with respect 
to the total weight of all clusters.
\label{fig:total:weight}}
\end{figure}

 Fig.\ \ref{fig:total:weight} displays the fraction
of the solutions contained in the largest cluster. For $\al<\al_{c}$ 
this value seems to increase with growing system size. At
$\al=\al_{c}$ it exhibits a minimum, while for $\al> \al_{c}$ it decreases
slightly with growing system size, but it is larger than the value found right
at $\al_{c}$. These results are also compatible with the analytical prediction 
\cite{Krzakala2007CSP}, which states that 
only for a range $\al\ge \al_{c}$ more than one cluster
is relevant in the thermodynamic limit. Nevertheless, we
 cannot deduce from the data,  since system sizes
we can reach are rather limited, whether for all values of  $\al<\al_{c}$
this growing fraction converges
to one or to a smaller value.

%
%

\begin{figure}
  \includegraphics[angle=-90,width=\linewidth]{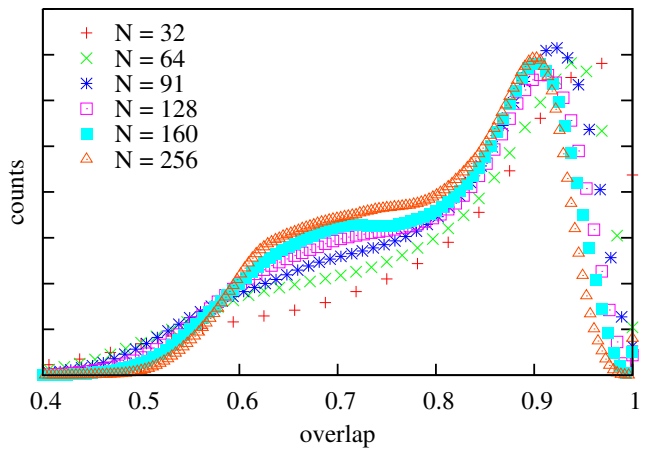}
  \caption{Overlap $\langle r\rangle$ of solutions in clusters found by Ballistic Networking, for $\al = 4.0$. For $\langle r\rangle < 0.4$ the curves are essentially zero.}
  \label{fig:bs_overlap1}
\end{figure}

Next, we have a closer look at the average structure of the
solution space. As mentioned in the discussion of the Ward matrices,
solutions belonging to the same cluster are more similar to each
other, i.~e., closer in terms of the Hamming distance, than pairs of
solutions which belong to different clusters. The cluster structure is
thus reflected in the set of all pairwise overlaps, where the overlap
$r_{ij}$ of two solutions $i$ and $j$ for which the Boolean variables
take the values given by $x_n^{(i)}$ and $x_n^{(j)}$ is defined as
$r_{ij} := \sum_{n=1}^N \delta(x_n^{(i)}, x_n^{(j)})/N$.  Fig.
\ref{fig:bs_overlap1} shows the overlap distribution for
$\al=4.0$ and several values of $N$.  For each system size at least
1000 instances have been processed with the algorithm described in
section \ref{sec:implementation} and 500 solutions have been generated
from each cluster survey.

Two peaks are visible. One peak is lying close to $\langle r\rangle =
1$ and due to the overlap of solutions belonging to the same
cluster. With larger system size it moves slightly to lower $\langle
r\rangle$ values and becomes sharper. The second peak at about
$\langle r\rangle = 0.7$ is not discernible for the smallest system
size, but only evolving with larger system sizes
and only visible weakly against an also growing background.
Note that a pure two-peak structure would correspond to  the picture of 
one-step broken replica symmetry (1-RSB) \cite{Mezard1984}.
Nevertheless, the result is not fully clear here, since in addition
to the peaks, there is also a contious part between the two peaks.
On the other hand, it is clear that the overlap converges to zero
for values of the overlap smaller than 0.5. This speaks against
a full-level RSB structure of the solutions space.
Note that we have found similar results for other values of 
the control parameter $\al_{c}<\al<\al_s$ (not shown).

\subsection{Freezing transition}


To complete the picture we also studied the freezing transition, which
as mentioned in section \ref{sec:cluster_phenomena} is defined as the
smallest \al above which all solutions belong to frozen clusters and
has been found to lie at $\al_f = 4.254$.

To check directly whether a cluster contains frozen
variables,  we need to generate and compare all solutions from this
cluster,  therefore cluster surveys do not help here.  Using exact
algorithms we find that for the system sizes we can reach, for all \al
near the SAT-UNSAT transition there are always frozen variables
in all clusters. This is probably due to too small systems
sizes.

\begin{figure}
  \includegraphics[width=0.9\linewidth]{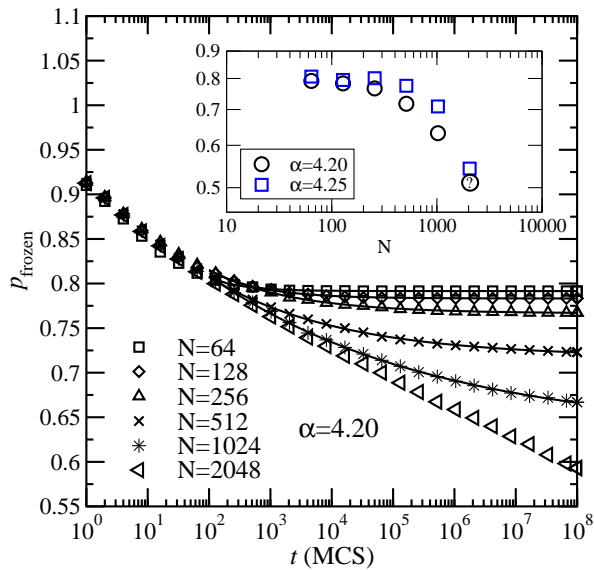}
  \caption{Fraction $p_{\rm frozen}(t)$ 
of never-flipped variables for the ``largest''
(i.~e. first detected) cluster during $T=0$ MC simulations for 
$\al=4.20$ and different system sizes $N$.
The solid lines indicate fits to functions of the form
$p_{\rm frozen}(t)=p_{\rm frozen}^{\infty}+at^{-\beta}$
Inset: Extrapolated values $p_{\rm frozen}^{\infty}$ as function of system
size $N$ in a double-logarithmic scale, for $\al=4.20$ and $\al=4.25$. 
The ``?'' marks a point where the extrapolation failed and an upper limit was 
estimated by eye.
  \label{fig:p:frozen}
  }
\end{figure}

Thus, we followed  a different approach. For each instance, 
taken at $\al=4.20$ and $4.25$ and for system sizes up
to $N=2048$, we generated
a solution using ASAT, which belongs with high probability to the largest
cluster. Then we performed a very long $T=0$ simulation starting from 
this solution, and measured
the fraction $p_{\rm frozen}$ of variables which have never flipped
while performing this random walk inside the solution cluster.
We extrapolated this fraction to a large number of MC steps, yielding
$p_{\rm frozen}^\infty$, see
Fig.\ \ref{fig:p:frozen}. 
With increasing system size, $p_{\rm frozen}^\infty$
seems to converge to zero, 
see inset of Fig.\ \ref{fig:p:frozen}. Hence for $\al=4.20$ and
$\al=4.25$ the largest clusters seems to contain no frozen variables
in the thermodynamic limit. This is
compatible with $\al_f = 4.254$, meaning that in the thermodynamic limit no 
frozen clusters occur below this value of \al.

\section{Conclusion}
In this work we have shown that stochastic local search algorithms cannot be expected to produce correctly weighted samples of the solution space of Satisfiability.
The same holds true for MCMCMC which is widely used to sample configuration spaces in many fields of application, when the SAT solutions are spread over several clusters.

A new type of algorithm has been presented and used for studying clustering phenomena in the solution landscape of the Satisfiability problem.
It is an improved version of the Ballistic Search algorithm which has been successfully used for studying spin-glasses.
Its guiding principle is to generate a survey of clusters of solutions represented by small sets of solutions rather than enumerating and clustering all solutions which is unfeasible already for moderate system sizes.
By using a different approach, Ballistic Networking,
in the reconstructing process of the cluster structure the efficiency of the Ballistic Search could be improved so that its performance becomes reasonably high when used on Satisfiability.
The method presented here is general enough to be suitable for many other problems. Of course, it would be natural to study Satisfiability for $K>3$ using Ballistic Networking, but the efficiency for ballistic path search seems to be still much lower than for the case of $K=3$ which sets very restricting limits on the system sizes which can be reached.
Nevertheless, the approach presented here should be useful for many
disordered systems like other types of combinatorial optimization problems.

In the case of Satisfiability, the range of low values of \al 
(where many solutions exist but belong to only one cluster)
 can be studied by MCMCMC. Furthermore, 
the case of high values of \al close
to the SAT-UNSAT transition (fewer solutions contained in 
several clusters) can be studied using exact enumeration of all solutions.
In contrast, the algorithm presented here allows to study the full 
satisfiable phase, but it is limited to moderate system sizes 
in the intermediate \al range. Nevertheless it is the only reliable 
method to generate unbiased samples in this regime.

Using the method described here the ensemble properties of Satisfiability with moderate system size could be studied and analytic predictions  about the cluster structure could be tested.
To this aim we first did a visual inspection of the cluster landscape using a graphical representation in terms of Ward distance matrices. These show the expected structural differences of the different phases of Satisfiability.
Furthermore we had a look at the complexity measure over the whole \al spectrum in the easy phase, and the overlap distribution of the solutions for particular values of \al.
Our findings are in good agreement with the theoretical predictions and previous numerical studies using other methods.

\begin{acknowledgments}
We have profited a lot from discussions with M. Alava,
J. Ardelius, E. Aurell, P. Kaski, 
F. Krzakala, M. M\'ezard, A. Montanari, P. Orponen, S. Seitz and 
L. Zdeborov\'a.
 This work was supported financially by the \emph{VolkswagenStiftung} 
within the
program ``Nachwuchsgruppen an Universit\"aten''.
We furthermore wish to acknowledge the allocation of
computer time  by the
  Gesellschaft für wissenschaftliche Datenverarbeitung mbH G\"ottingen, by the
  Institute of Theo\-reti\-cal Physics (University of G\"ottingen), and
by  the 
Cluster for Scientific Computing ``GOLEM'' (University of Oldenburg).
\end{acknowledgments}

\bibliography{artikel}


%
%

\end{document}